\ifdefined\pdfminorversion\pdfminorversion=7\relax\fi
\PassOptionsToPackage{pdfversion=1.7}{hyperref}

\documentclass[sigplan,10pt,nonacm]{acmart}

\renewcommand\footnotetextcopyrightpermission[1]{}
\AtBeginDocument{}

\usepackage{amsmath,amssymb,amsfonts}
\usepackage{graphicx}
\usepackage{textcomp}
\usepackage{xcolor}
\usepackage{booktabs}
\usepackage{stfloats}
\usepackage{multirow}
\usepackage{array}
\usepackage{url}
\usepackage{marvosym}
\usepackage[most]{tcolorbox}

\definecolor{findingblue}{HTML}{385375}
\definecolor{findingbackground}{HTML}{ECF2F8}
\newtcolorbox[auto counter]{finding}[1][]{%
  enhanced, breakable=false,
  colback=findingbackground, colframe=findingblue,
  colbacktitle=findingblue, coltitle=white,
  boxrule=0.6pt, arc=3pt, boxsep=0pt,
  left=7pt, right=7pt, top=4pt, bottom=5pt,
  toptitle=3pt, bottomtitle=3pt,
  before skip=4pt, after skip=6pt,
  fonttitle=\bfseries,
  fontupper=\bfseries,
  title={Finding~\thetcbcounter},
  #1}

\newcommand{\sysname}{\textsc{NORmal}}
\newcommand{\circled}[1]{\textcircled{\scriptsize #1}}
\newcommand{\corrmail}{\texorpdfstring{\Letter}{}}

\begin{document}

\title{\sysname: Architecting 3D Ferroelectric NOR Flash as Main Memory}
\subtitle{From Co-Fabricated BEOL Arrays to 200B-Class Multi-Agent LLMs in a Smartphone}

\author{%
\sffamily
Zhuoran Li\textsuperscript{1} \quad
Zhuohang Bian\textsuperscript{1} \quad
Yibo Zhao\textsuperscript{1} \quad
Xuxiao Yang\textsuperscript{1}  \quad
Youwei Zhuo\textsuperscript{1,\corrmail}%
}
\affiliation{%
  \institution{\vspace{2pt}\textsuperscript{1}Peking University, Beijing, China\\
  \{zhuoranli, zhbian26, yibozhao, valinor\}@stu.pku.edu.cn
  \hspace{0.8em} youwei@pku.edu.cn}%
  \country{}%
}

\renewcommand{\shortauthors}{Zhuoran Li et al.}

\begin{abstract}
\begin{table*}[!b]
\centering
\caption{Mobile memory and flash tiers on one 96-bit interface.}
\label{tab:device_comparison}
\renewcommand{\arraystretch}{1.15}
\setlength{\tabcolsep}{5pt}
\resizebox{\textwidth}{!}{%
\begin{tabular}{lcccccccc}
\toprule
\multirow{2}{*}{\textbf{Memory Tier}} & \multirow{2}{*}{\textbf{\shortstack{Capacity per\\96-bit Interface}}} & \multirow{2}{*}{\textbf{\shortstack{Host Interface \&\\Peak Read Bandwidth}}} & \multicolumn{2}{c}{\textbf{Latency}} & \multicolumn{2}{c}{\textbf{Energy per Bit}} & \multirow{2}{*}{\textbf{\shortstack{Minimal\\Access Unit}}} & \multirow{2}{*}{\textbf{\shortstack{Endurance\\(P/E Cycles)}}} \\
\noalign{\vskip -3pt}
\cmidrule(lr){4-5}\cmidrule(lr){6-7}
\noalign{\vskip -3pt}
 & & & \textbf{Read} & \textbf{Write} & \textbf{Read} & \textbf{Write} & & \\
\midrule
\textbf{LPDDR6 DRAM}~\cite{jedec2025lpddr6} & 12--24~GB & 96-b LPDDR6 (153.6~GB/s) & 30~ns & 15~ns & 1.5~pJ & 1.5~pJ & 32~B & $>10^{15}$ \\
\textbf{UFS 4.0/5.0 (3D NAND)}~\cite{powerinfer2} & 256--1024~GB & Serial M-PHY (4.2--10~GB/s) & 40~\textmu s & 500~\textmu s & $>$10~pJ & $>$100~pJ & 4~KiB & $10^3$--$10^4$ \\
\textbf{HBF (zNAND-O)}~\cite{samsung2025znando,flashaccel2026} & 64--128~GB (4-Hi) & Custom Wide I/O + Substrate (200~GB/s) & 4~\textmu s & 100~\textmu s & 8~pJ & 100~pJ & 4~KiB & $10^5$ \\
\textbf{\sysname{} (Ours)} & \textbf{\shortstack{MLC: 49--197~GB\\SLC: 4--24~GB}} & \textbf{Shared LPDDR6 (153.6~GB/s)} & \textbf{30~ns} & \textbf{100~ns (SLC)} & \textbf{0.45~pJ} & \textbf{0.45~pJ} & \textbf{96~B / 64~B} & $\mathbf{10^8 / >\!10^{11}}$ \\
\bottomrule
\end{tabular}%
}
\vspace{2pt}
\par\noindent\parbox{\textwidth}{\centering\scriptsize \sysname{} MLC is written only at model installation; its write numbers refer to SLC. \\HBF additionally requires a redesigned SoC, TSVs, and a custom SRAM base die at $3\times$ UFS cost per byte.}
\vspace{-4pt}
\end{table*}%
Smartphone agents increasingly work as teams, in which a large planner directs parallel workers across long sessions. Running such a team on the device requires 200B-class models and contexts that exceed phone DRAM many times over, yet existing tiers treat flash as storage: UFS starves per-token weight streaming, and mobile High-Bandwidth Flash (HBF) needs a redesigned SoC and cannot absorb continuous KV writes. We show that a non-volatile memory can join an unmodified LPDDR6 bus as a plain rank only if its access latency fits the DRAM timing window, a condition no NAND tier meets.
We present \sysname{}, to our knowledge \textbf{the first main-memory architecture for 3D ferroelectric NOR flash.}
We have fabricated and experimentally characterized back-end-of-line (BEOL) 3D FeNOR arrays that co-integrate dense 3-bit MLC and fast, high-endurance in-place-writable SLC with sub-7.5~ns sensing.
\sysname{} turns these co-fabricated arrays into a byte-addressable, deterministic-timing load/store rank on an unmodified LPDDR6 interface through an under-array DRAM-emulation bridge, staggered bank reads with zero-cycle weight swizzling, and posted SLC writes.
Across this rank and DRAM, \sysname{} places agent state by mutability and lifetime rather than by hotness, serving 200B-class models and persistent planner contexts in place while dynamically partitioning subchannels across agent phases.
In serving-level simulation on production agent traces, \sysname{} is $8.9\times$--$28.0\times$ faster than UFS offloading, within 0.4\% of an unlimited-DRAM ideal, and $2.8\times$--$6.3\times$ faster than HBF on sessions beyond 256K tokens, with $8.1\times$--$17.8\times$ less memory energy per token.
\end{abstract}

\maketitle
\thispagestyle{plain}
\pagestyle{plain}

\section{Introduction}
\label{sec:intro}

Smartphone agents now complete tasks rather than answer questions. They read screens and sensors, hear from you, operate apps, and act on messages and credentials, so they must run on the device for privacy.
They also increasingly work as teams~\cite{claudecodesubagents,kimik3cluster}. A \emph{planner} holds the long-running session and dispatches subtasks to parallel \emph{workers} that run a smaller model, return a result, and exit.
In the production traces we replay~\cite{wekatraces2026}, a planner keeps up to ten workers in flight over sessions of up to 1,970 turns.
Reliable planning needs large Mixture-of-Experts (MoE) models~\cite{deepseekv3,hybrimoe2025}, whose sparse activation keeps compute tractable on a mobile NPU.
Memory is the bottleneck. A 200B-class planner alone exceeds usable DRAM by nearly $20\times$, since the operating system and background apps already hold over half of a phone's memory~\cite{powerinfer2}.

Our characterization of these traces shows that the team's memory demands follow its structure, and that each demand breaks the conventional split between DRAM and flash storage.
\emph{(i)}~Both models must stay resident and be read at memory speed.
DRAM holds only a few percent of a large MoE model, and because routers shift active experts across tokens, caching and prefetching still leave gigabytes of weights to fetch per token~\cite{hybrimoe2025,qwen38expertmap2026}. Swapping a 167~GB planner over UFS instead costs 15~s whenever control passes between the planner and its workers.
\emph{(ii)}~The planner's context outgrows DRAM and is written at every token.
Long multi-turn histories, which on phones also accumulate screenshots and audio~\cite{claudecomputeruse,mobileworld2025}, push the planner's KV well past usable DRAM.
\emph{(iii)}~The team's state splits by lifetime, and its traffic shifts by phase.
Planner KV persists for the entire session, worker KV is discarded as soon as each worker returns, and bus demand alternates between planner decode and worker execution.

Existing mobile tiers cannot meet these demands because they treat flash as block storage rather than memory.
UFS NAND provides ample capacity, but its narrow serial link and microsecond page reads throttle per-token weight streaming~\cite{powerinfer2,llminaflash2024}.
Mobile High-Bandwidth Flash (HBF), such as Samsung's zNAND-O~\cite{samsung2025znando,flashaccel2026}, widens the read link using 3D-stacked NAND, TSVs, a custom SRAM base die, a silicon interposer, and a redesigned SoC interface.
However, HBF remains page-granular NAND. Coarse pages amplify fine-grained MoE fetches~\cite{hbfsucks2026}, and high write energy and limited endurance rule out per-token KV appends within a phone's power and lifetime budgets.
Once planner KV outgrows DRAM, HBF must spill it to UFS or recompute it.

Can flash instead serve as \emph{main memory} on the bus the phone already has?
An unmodified LPDDR6 controller assumes a fixed first-access latency and a fixed burst rate.
Bank-level parallelism can supply the burst rate, but first-access latency is a device property that architecture cannot fix, and NAND misses it by more than $100\times$.

We have fabricated and experimentally characterized back-end-of-line (BEOL) 3D ferroelectric NOR (FeNOR) flash arrays that co-integrate dual-gate SLC and gate-all-around 3-bit MLC above CMOS logic (Fig.~\ref{fig:fenor_device}).
Our arrays combine three properties that turn flash into a plain rank on the existing LPDDR6 bus.
\emph{(i)}~Measured sub-7.5~ns array sensing keeps a full read, including CMOS-under-Array (CuA) peripherals and the LPDDR6 PHY, inside the 25--35~ns DRAM timing window.
\emph{(ii)}~A $\le\!400^\circ$C BEOL flow stacks 16--32 monolithic array layers directly above CMOS logic and sense amplifiers at 0.45~pJ/bit.
\emph{(iii)}~One flow co-fabricates gate-all-around 3-bit MLC (with a measured 4.6~V eight-state memory window) for dense weights and dual-gate SLC (with $>10^8$ ON/OFF current ratio and $>10^{11}$ endurance) for 100~ns in-place writes on the same die.
The tradeoff is explicit.
FeNOR shares the bus's 153.6~GB/s instead of adding a link as HBF does, and in return it provides writable capacity, sub-picojoule access, and unmodified SoC hardware.

\textbf{We present \sysname{}, to our knowledge the first main-memory architecture for 3D ferroelectric NOR flash.}
Rather than acting as read-only execute-in-place ROM, \sysname{} exposes co-fabricated MLC and SLC arrays as a byte-addressable, deterministic-timing load/store rank behind an unmodified LPDDR6 controller.
In the CuA logic layer, staggered bank interleaving and a zero-cycle swizzle latch turn banks that sense at one third of the subchannel rate and store 3-bit cells into a continuous 4-bit weight stream at full bus speed.
A posted-write buffer with pending-read forwarding hides 100~ns in-place SLC writes behind 15~ns DRAM write timing.
Above this rank, \sysname{} maps the agent hierarchy onto the memory hierarchy, placing data by mutability and lifetime rather than by hotness.
Both models are immutable and always needed, so they live in MLC and are never swapped.
The planner's context is persistent and append-only, so it lives in in-place-writable SLC.
Worker contexts are short-lived, so they stay in DRAM and are freed as each worker returns.
Hotness caching is unnecessary because, on a shared bus, a FeNOR read takes the same wire time as a DRAM read, so the NPU reads weights and planner KV in place.
To follow the team's phases, adaptive subchannel allocation re-divides the shared channel between FeNOR and DRAM at each phase boundary.

We evaluate \sysname{} with a full-stack serving simulator that replays single-agent and multi-agent sessions. \sysname{} is $8.9\times$--$28.0\times$ faster than UFS~5.0 offloading ($21.4\times$ for the agent team) and reaches over 99.6\% of the throughput of an unlimited-DRAM ideal. HBF's dedicated link keeps it $1.10\times$--$1.73\times$ faster on light workloads, but without UFS spilling HBF cannot run up to 40\% of sessions, and on sessions longer than 256K tokens \sysname{} finishes $2.8\times$--$6.3\times$ faster. \sysname{} also uses $8.1\times$--$17.8\times$ less memory energy per token than HBF, and SLC endurance removes write wear-out as a lifetime limit. We make four contributions:
\begin{itemize}
    \item \textbf{Characterization of on-device agent teams (\S~\ref{sec:background}).} On production traces and six models, we show that a planner--worker team needs resident models read at memory speed, writable capacity for planner KV beyond DRAM, and lifetime-aware placement, and that DRAM expert caching, UFS offloading, and HBF each fail at least one of these needs.
    \item \textbf{Fabricated 3D FeNOR as a plain LPDDR6 rank (\S~\ref{sec:rank}).} We separate a device-level latency bound from an architecture-level bandwidth bound for any memory behind an unmodified LPDDR6 controller, and show with fabricated and experimentally characterized BEOL 3D FeNOR arrays that co-integrated 3-bit MLC and in-place-writable SLC meet the latency bound.
    \item \textbf{The \sysname{} architecture (\S~\ref{sec:architecture}).} We present, to our knowledge, the first main-memory architecture for 3D ferroelectric NOR flash. A dual-tier MLC/SLC rank with a CuA DRAM-emulation bridge, zero-cycle weight swizzling, and posted SLC writes pairs with an inference-in-place runtime to map the agent hierarchy onto the memory hierarchy.
    \item \textbf{Serving-level evaluation (\S~\ref{sec:evaluation}).} Against UFS~5.0 offloading, mobile HBF, and an unlimited-DRAM ideal, we report both where HBF's extra bandwidth wins and where \sysname{}'s writable main-memory capacity wins.
\end{itemize}

\section{Background and Characterization}
\label{sec:background}

\subsection{Mobile Memories}
\label{sec:tiers}

A flagship smartphone SoC reaches DRAM through one 96-bit LPDDR6 interface organized as eight independent 12-bit subchannels, providing up to 153.6~GB/s~\cite{jedec2025lpddr6}. A typical device ships with 16~GB of DRAM, but much less is available to any one application. Even on a freshly managed phone, PowerInfer-2 measures that a single app can occupy at most 11~GB of a 16~GB device~\cite{powerinfer2}. In ordinary use, vendor Android builds keep many background apps and services resident. We therefore model an inference budget of 8~GiB on a 16~GB device, after the operating system, background services, and a few-MiB runtime reserve (Sec.~\ref{sec:methodology}). Table~\ref{tab:device_comparison} compares the tiers that can extend this capacity.

\noindent\textbf{UFS NAND.} Phones store model weights in UFS~4.0/5.0 NAND flash, which provides hundreds of gigabytes behind a serial M-PHY link of 4.2--10~GB/s. Reads take tens of microseconds and fetch whole 4~KiB pages. Existing on-device systems stream the active parameters from UFS into DRAM and cache frequently used parameters in DRAM~\cite{powerinfer2,llminaflash2024,hybrimoe2025,d2moe2025}.

\noindent\textbf{Mobile HBF.} High-Bandwidth Flash stacks NAND dies on a logic base die and connects them to the SoC through a dedicated wide-I/O link on a custom interposer~\cite{ocphbf2025,hbfsucks2026,flashaccel2026,flint2026,li2026hbfsim,Ha2026H3HBF,Hu2026HBFSim,Kim2026HBFRoadmap,Son2026ExploringHBF,Yin2026HBFApplications,Zhong2026HBFlex,Petrucci2026HBFAllYouNeed,hsu2026havenhighbandwidthflashaugmented,Yu2026CharacterizingHBF}. Samsung's zNAND-O targets mobile devices~\cite{samsung2025znando}. It stacks four SLC NAND dies (4-Hi) for low read latency and integrates several MiB of SRAM in the base die to hide microsecond array latency. The link provides up to 200~GB/s. Passive cooling limits a stack to 4-Hi, which gives 64--128~GB per stack, so up to 171.2~GB of co-resident weights in our multi-agent workload needs 2--3 stacks. Reads cost 8~pJ/bit and writes cost 100~pJ/bit, endurance is $10^5$ P/E cycles, and the TSVs, base die, organic substrate, and SoC redesign raise cost per byte to $3\times$ that of UFS. Page-granular reads amplify fine-grained accesses, by up to $13.6\times$ on 320~B rows~\cite{hbfsucks2026}.

\begin{figure}
    \centering
    \includegraphics[width=1.03\linewidth]{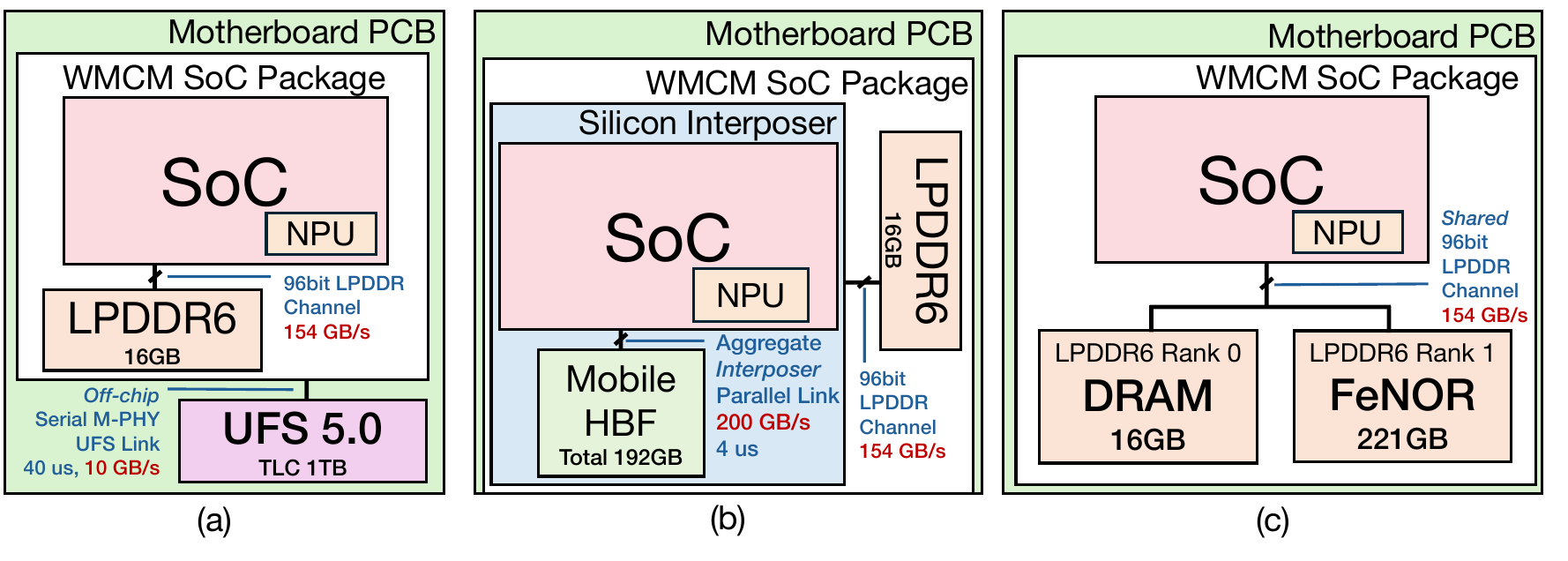}
    \caption{Mobile memory layouts: (a)~off-chip UFS~5.0, (b)~interposer Mobile HBF, and (c)~\sysname{} on shared LPDDR6.}
    \label{fig:layouts}
\end{figure}

\subsection{On-Device Agent Workloads}
\label{sec:workloads}

\noindent\textbf{Models.} We study six models (Table~\ref{tab:workloads}): two dense models and four MoE models from 35B to 314B parameters. The MoE models activate 3B--17B parameters per token, and a 200~TOPS INT4 NPU has ample compute for single-user decode. The 100B-class MoE models have weight footprints of 94.3--166.5~GB at 4 bits,  exceeding usable phone DRAM by up to $19\times$.

\noindent\textbf{Agent structure.} Personal agents decompose tasks into a \emph{planner--worker} team. A \emph{planner} running a large model holds the session, plans the task, and dispatches subtasks to \emph{workers} that run a smaller, faster model in parallel. In the production agent traces we replay~\cite{wekatraces2026}, the planner can have up to ten workers in flight at once. Our default team pairs a Qwen3.8-Flash-Next planner with Qwen3.5-9B workers. We choose same-family models so that the planner and its workers share a tokenizer and prompt format, which keeps context consistent as it passes between them and keeps the team comparable with the single-model Qwen workloads. Both models must stay resident. Reloading the Flash-Next planner over a 10.8~GB/s UFS link would take about 9~s. 

 Mobile agents add multimodal observations that increase memory pressure. A screenshot costs roughly 1,000--1,800 input tokens~\cite{claudecomputeruse}, and audio adds tokens in proportion to its duration~\cite{osworld2024,uitars2025}. Production phone agents such as the Doubao phone assistant run tasks of 30 or more steps that switch between foreground and background execution, and mobile benchmarks already average 27.8 steps per task~\cite{mobileworld2025}. Keeping only a few recent observations would discard state these tasks need. 

\subsection{Characterization}
\label{sec:characterization}
\label{sec:data_classes}
Several works coordinate scheduling and KV-cache management using agent dependencies and tool-call intervals~\cite{bian2026tokencakekvcachecentricservingframework,bian2026tokendancescalingmultiagentllm}.
We replay the traces against the six models with the simulator described in Sec.~\ref{sec:methodology}. Four findings follow, and each yields a requirement for the memory system.

\noindent\textbf{Weight streaming and expert caching.} Existing systems cache frequently used experts in DRAM and prefetch predicted ones~\cite{powerinfer2,hybrimoe2025,d2moe2025}. Training balances expert load only globally, so individual requests and domains still favor subsets of experts~\cite{demons2025,moeinfinity2024,localrouting2026}. Hidden-state similarity also lets systems predict next-layer experts with about 90\% accuracy~\cite{adapmoe2024,hobbit2024}. Phone DRAM limits what this locality can buy. Even if all 8~GiB held experts, it would cover under 6\% of a 151--167~GB model, and  caches hit only 30--53\% of accesses even at a 25\% cache~\cite{hybrimoe2025,qwen38expertmap2026}\footnote{\url{https://huggingface.co/datasets/tcclaviger/Qwen3.8-Flash-Next-expert-activation-map}}. Prediction hides fetch latency but not fetch volume, because every miss still crosses the UFS link. At 4 bits, the 3B--17B active parameters take 1.5--8.5~GB per token. Without cache hits, a 10.8~GB/s UFS link therefore caps decode at 1.3--7.2~tok/s, against 18--102~tok/s at full LPDDR6 bandwidth, and a cache with hit rate $h$ raises this cap only by a factor of $1/(1-h)$. The cache also competes with KV for the same DRAM (Finding~\ref{find:kv_capacity}).

\begin{finding}[label=find:weights]
Phone DRAM is too small for expert caching to remove the UFS bottleneck. \\
\textit{Requirement:} the capacity tier itself must deliver weights at memory bandwidth and latency.
\end{finding}

\noindent\textbf{KV growth.} Using the per-token KV sizes in Table~\ref{tab:workloads}, a single Qwen3.8-27B session of the traces' mean prompt length (223K tokens) needs 7.2~GiB of KV, about the entire usable DRAM. The longest session (962K tokens) needs 31~GiB. With one planner and four concurrent workers at 200K context, KV reaches  $1.8\times$--$2.4\times$ the usable DRAM. This KV is appended at every decoded token and re-read at every attention step.

\begin{finding}[label=find:kv_capacity]
KV caches outgrow DRAM within a single session.
\textit{Requirement:} the system needs a writable tier, beyond DRAM, with memory-speed reads and enough endurance for continuous appends.
\end{finding}

\noindent\textbf{KV lifecycles.} The two kinds of KV behave differently. The planner's KV is append-only and persists across the entire multi-turn session, so it grows large. Each worker's KV lives only for one subtask and is discarded when the worker returns its result.

\begin{finding}[label=find:lifecycle]
Planner KV persists for the whole session, while worker KV lives only for one subtask.\\
\textit{Requirement:} persistent planner KV belongs in a writable capacity tier, while ephemeral worker KV belongs in DRAM so that it can be freed immediately.
\end{finding}

\noindent\textbf{Phase-dependent traffic.} During planner decode, traffic is dominated by weight and planner KV reads. During worker execution and prefill, DRAM traffic (worker KV, activations, and recurrent state) grows. Any design that serves both kinds of traffic from one channel faces a demand mix that changes at phase boundaries; Sec.~\ref{sec:eval_ablation} shows that fixed splits lose 9\%--87\% of throughput.

\begin{finding}[label=find:phase]
The mix of capacity-tier and DRAM traffic shifts at agent phase boundaries.\\
\textit{Requirement:} if capacity-tier and DRAM traffic share a channel, the split must follow the phase.
\end{finding}

\subsection{Limitations of Existing Memory Tiers}
\label{sec:tiers_fall_short}

\noindent\textbf{DRAM caching.} DRAM provides the bandwidth, latency, and writability that Findings~\ref{find:weights}--\ref{find:phase} require, but not the capacity. Up to 171~GB of co-resident weights is an order of magnitude beyond phone DRAM, and Finding~\ref{find:weights} shows that caching a subset does not substitute for capacity.

\noindent\textbf{UFS offloading} fails the weight requirement of Finding~\ref{find:weights}, as its serial link caps decode at a few tokens per second. It also fails the KV requirement of Finding~\ref{find:kv_capacity}. Spilled KV must cross the same link, and the NAND endures only $10^3$--$10^4$ P/E cycles.

\noindent\textbf{Mobile HBF} meets the weight requirement of Finding~\ref{find:weights} with a dedicated link, although 4~KiB pages waste part of it on fine-grained reads. It fails the KV requirement of Finding~\ref{find:kv_capacity} on a phone. Datacenter HBF systems buffer and coalesce append-only KV writes~\cite{flashaccel2026} under active cooling, but coalescing reduces write amplification without lowering the intrinsic 100~pJ/bit NAND programming energy. Under a phone's passive thermal budget, sustained GB/s KV writes during long-context prefill or DRAM-overflow eviction would consume several watts in flash programming alone, forcing mobile HBF to operate as a read-only weight tier. Any HBF capacity left over after storing weights therefore sits idle. Even if thermal limits were ignored, $10^5$ P/E endurance would still wear out within months to a few years (Sec.~\ref{sec:eval_write}). KV must stay in DRAM and  be evicted once it exceeds DRAM (Finding~\ref{find:kv_capacity}). HBF also requires a redesigned SoC. The extra link is a real advantage while KV fits in DRAM.

No existing tier offers memory-speed, writable capacity on the interface the SoC already has. The next section shows when a memory can provide it.

\section{Flash as a Plain LPDDR6 Rank}
\label{sec:rank}


\subsection{The Rank Condition}
\label{sec:rank_condition}

An unmodified LPDDR6 controller makes two timing assumptions about every rank~\cite{jedec2025lpddr6}. After a row activate (\texttt{ACT}), the first column read returns within the programmed $t_{RCD}+RL$. After that, 32~B bursts stream every $t_{CCD}=1.67$~ns on each 12-bit subchannel (19.2~GB/s). A non-volatile memory can therefore serve read bursts as a standard rank behind an unmodified controller only if
\begin{align}
t_{\text{access}} &\le t_{RCD} + RL, \label{eq:rank_latency}\\
N_{\text{bank}} \cdot \frac{U_{\text{row}}}{t_{\text{bank}}} &\ge \frac{32~\text{B}}{t_{CCD}}, \label{eq:rank_bw}
\end{align}
where $U_{\text{row}}$ is the data one bank senses per cycle $t_{\text{bank}}$, and $N_{\text{bank}}$ is the number of banks whose cycles can overlap. Eq.~\ref{eq:rank_latency} is a device-level requirement that no amount of architecture can relax. Eq.~\ref{eq:rank_bw} is an architecture-level requirement that adding banks can satisfy.

\noindent\textbf{Prior non-volatile ranks.} NAND flash misses Eq.~\ref{eq:rank_latency} by more than $100\times$ (4--50~\textmu s), which is why UFS and HBF remain block storage behind their own interfaces. Lincoln~\cite{lincoln2025} shares LPDDR channels between DRAM and NAND flash ranks to run 50--100B models on consumer devices. Because NAND cannot meet Eq.~\ref{eq:rank_latency}, Lincoln stages data in near-PHY SRAM using explicit commands and polling and computes generation-phase dot products near the flash, while KV stays in DRAM. Cambricon-LLM~\cite{cambriconllm2024} similarly pairs an NPU chiplet with compute-enabled NAND and keeps KV in DRAM. Phase-change memory~\cite{lee2009pcm,qureshi2009pcm} and Optane persistent memory, with $\sim$300~ns reads~\cite{izraelevitz2019optane}, also miss Eq.~\ref{eq:rank_latency}, and Optane used a non-deterministic DDR-T protocol with a modified controller. Ferroelectric 1T1C capacitor RAM (FeDRAM, such as Micron's 2-layer, 32~Gb NVDRAM~\cite{nvdram2023}) meets DRAM timing on LPDDR5, but it is a destructive-read 1T1C RAM rather than flash. Each read requires a write-back restore, cells store only 1~bit, and density stays at the DRAM scale (4~GB/die) rather than the 100~GB scale of multi-layer, multi-bit 3D flash. Sharing a mobile bus with NVM has precedent. JEDEC LPDDR2-NVM~\cite{jedec2010lpddr2} let PCM and SDRAM share command and data pins with separate chip selects, and it defined NVM-specific initialization, buffering, and program operations.  SHyLA~\cite{shyla2026} places weights and KV across 3D-stacked PCM and DRAM, but in a custom cloud accelerator rather than on a phone's LPDDR bus.

\subsection{3D FeNOR Meets the Latency Bound}
\label{sec:device_bg}

\noindent\textbf{Conventional 3D NOR.} NOR flash reads randomly in under 100~ns and is byte-addressable, so it can serve as execute-in-place \emph{memory} rather than block storage; LPDDR4-interfaced NOR products already exist~\cite{semper2023,vlsi3dnor2026}. Conventional floating-gate or charge-trap 3D NOR, however, lacks the multi-bit density to hold 150--200~GB of weights. It also needs front-end thermal processing above $800^\circ$C, which prevents stacking over CMOS logic, and its high-voltage charge pumps consume 22.5~pJ/bit when programming.

\noindent\textbf{Our fabricated 3D FeNOR.} We fabricate 3D ferroelectric NOR (FeNOR) arrays in a back-end-of-line (BEOL) flow at or below $400^\circ$C, using atomic-layer-deposited InGaZnO$_4$ (IGZO) channels and Hf$_{0.5}$Zr$_{0.5}$O$_2$ (HZO) ferroelectric gate dielectrics~\cite{igzofefet2025}. One flow co-integrates two cell structures: dual-gate (DG) single-level cells (SLC, 1~bit per cell) and gate-all-around (GAA) multi-level cells (MLC, 3~bits per cell), both at a 150~nm pitch. The SLC array shows an ON/OFF read-current ratio above $10^8$ (2~\textmu A versus 10~fA. The MLC array opens a 4.6~V memory window that separates eight threshold states. Cells written to the same state respond uniformly across both arrays.

\begin{figure}
    \centering
    \includegraphics[width=\linewidth]{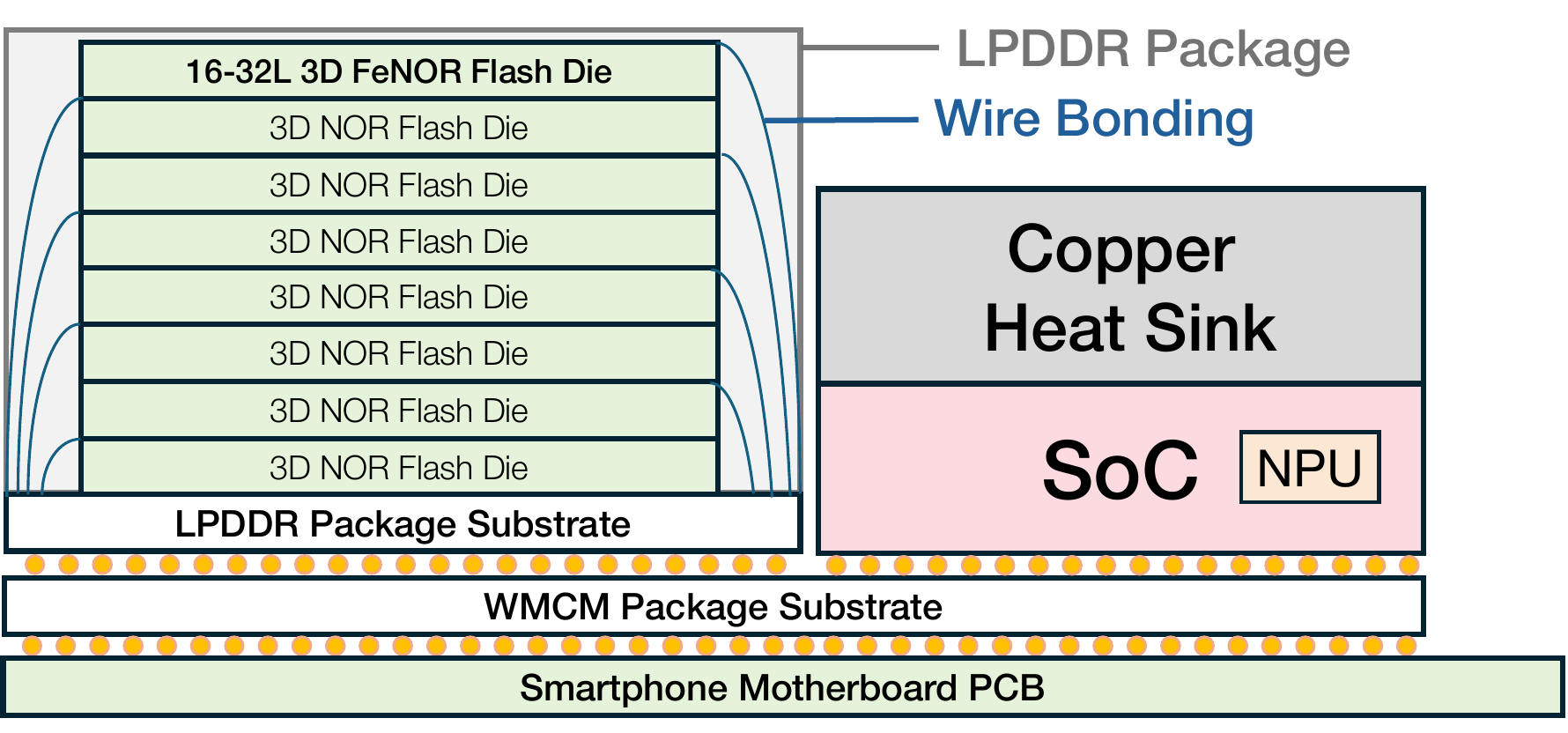}
    \caption{\sysname{} WMCM package placing the 8-die 3D FeNOR stack beside the SoC and copper heat sink.}
    \label{fig:package}
\end{figure}

\noindent\textbf{Properties for an LPDDR rank.} Together with CMOS-under-Array (CuA) integration, our 3D FeNOR arrays provide three properties that satisfy the rank latency bound and Findings~\ref{find:weights}--\ref{find:lifecycle} (Table~\ref{tab:device_comparison}).
\begin{enumerate}
    \item \textbf{Nanosecond sensing (Eq.~\ref{eq:rank_latency} and Finding~\ref{find:weights}).} Measured cell and array characterization~\cite{igzofefet2025,pcaa2025} shows that both cell types sense in under 7.5~ns. Array sensing, CuA peripheral settling, and the LPDDR6 PHY total 25--35~ns, fitting within $t_{RCD}+RL$. Unlike NAND flash, 1T1C FeDRAM, or conventional 1-bit NOR, 3D FeNOR combines multi-bit 3D flash density with sub-35~ns non-destructive reads inside the latency window of a plain LPDDR6 rank. A single bank, however, senses 96~B per 15~ns (6.4~GB/s), or one third of the 19.2~GB/s subchannel rate in Eq.~\ref{eq:rank_bw}; Sec.~\ref{sec:read_path} closes this gap with bank interleaving.
    \item \textbf{Multi-layer BEOL density and sub-picojoule access (Finding~\ref{find:weights}).} The $\le\!400^\circ$C BEOL thermal budget projects to 16--32 monolithic array layers per die at 85\% CuA area efficiency without degrading underlying CMOS. Because sense amplifiers, write drivers, and the LPDDR6 bridge logic sit directly beneath the vertical bitlines, measured array sensing and calibrated SPICE/CACTI modeling of the CuA peripherals and PHY keep die-level read and SLC write energy at 0.45~pJ/bit, $18\times$ below HBF read energy.
    \item \textbf{Co-fabricated MLC and SLC modes on one die (Findings~\ref{find:kv_capacity} and~\ref{find:lifecycle}).} The two cell structures trade write cost against density differently. GAA MLC stores $3\times$ the bits per cell, programs rows via a 100~\textmu s erase-and-verify, endures $10^8$ P/E cycles, and is more sensitive to thermal drift across its eight states, so it suits immutable weights. DG SLC switches a single cell in place in 100~ns without block erase, endures more than $10^{11}$ P/E cycles, and is insensitive to drift, so it suits persistent KV appends.
\end{enumerate}

\subsection{LPDDR6 Support for a Flash Rank}
\label{sec:lpddr6_hooks}

\sysname{} relies on three standard LPDDR6 features~\cite{jedec2025lpddr6}. First, mode registers (\texttt{MR}) set timing per rank, so Rank~1 (FeNOR) can advertise a 25--35~ns $t_{RCD}+RL$ and the deterministic 1.67~ns $t_{CCD}$ independently of Rank~0 (DRAM). Second, the eight independent 12-bit subchannels let DRAM and FeNOR traffic be divided per subchannel. Third, the 24-UI burst frame (\texttt{BL24}, 288 bits, carrying a 32~B payload and 32 metadata bits) on a 12-bit subchannel aligns with 12~bits, the least common multiple of 3-bit cells and 4-bit weights. In addition, on-the-fly 32~B/64~B burst control and dynamic non-target on-die termination (\texttt{NT-ODT}) support 96~B MLC weight fetches, 64~B SLC KV accesses, and fast rank turnaround.

\noindent\textbf{Unmodified SoC hardware.} The SoC's memory controller, PHY, and NPU are unchanged. All NOR-specific logic (the DRAM-emulation bridge, swizzle latch, posted-write buffer, and scrubber) sits in the CuA layer beneath the array. The changes are software, relying on standard controller support for per-rank timing registers and extended row addressing. Boot firmware programs Rank~1's mode registers and exposes the 221~GB rank via upper row-address bits that the CuA bridge decodes into die, layer, and bank selects. Because the controller already maps the eight 12-bit subchannels to distinct physical address slices with independent per-subchannel queues, the operating system maps the MLC and SLC regions as physical address ranges, and the NPU runtime steers bandwidth across subchannels purely in software by issuing DMA descriptors to the target subchannel address slices (Sec.~\ref{sec:runtime}).

\subsection{Remaining Challenges}
\label{sec:remaining_gaps}

FeNOR physics and the LPDDR6 features make a memory-mapped flash rank feasible. Three gaps remain between the device and a working rank that meets the requirements of Findings~\ref{find:weights}--\ref{find:phase} (Sec.~\ref{sec:characterization}).

\noindent\textbf{Challenge 1: Read path (Finding~\ref{find:weights}).} A single bank delivers one third of the subchannel rate (Eq.~\ref{eq:rank_bw}), and 3-bit MLC cells must deliver byte-aligned 4-bit weights. Padding each weight to two cells wastes 33\% of capacity, and unpacking in software stalls the NPU vector pipeline. 

\noindent\textbf{Challenge 2: Write path and placement (Findings~\ref{find:kv_capacity} and~\ref{find:lifecycle}).} A 100~\textmu s MLC erase-and-verify cannot absorb runtime KV appends. Even a 100~ns SLC write exceeds the DRAM write timing the controller expects ($t_{WR}\approx 15$~ns). Stacking 3-bit MLC directly on an 85$^\circ$C SoC in package-on-package (PoP) accelerates drift across its eight states. Data placement must also follow the lifecycle split between persistent planner KV and ephemeral worker KV. 

\noindent\textbf{Challenge 3: Shared bus (Finding~\ref{find:phase}).} Multi-gigabyte FeNOR streams of weights and planner KV share one 96-bit channel with latency-critical DRAM traffic (OS, activations, and worker KV). Static subchannel splits strand bandwidth as multi-agent execution moves between phases. 

\section{Overview}
\label{sec:overview}
\begin{figure*}[t]
\centering
\includegraphics[width=\textwidth]{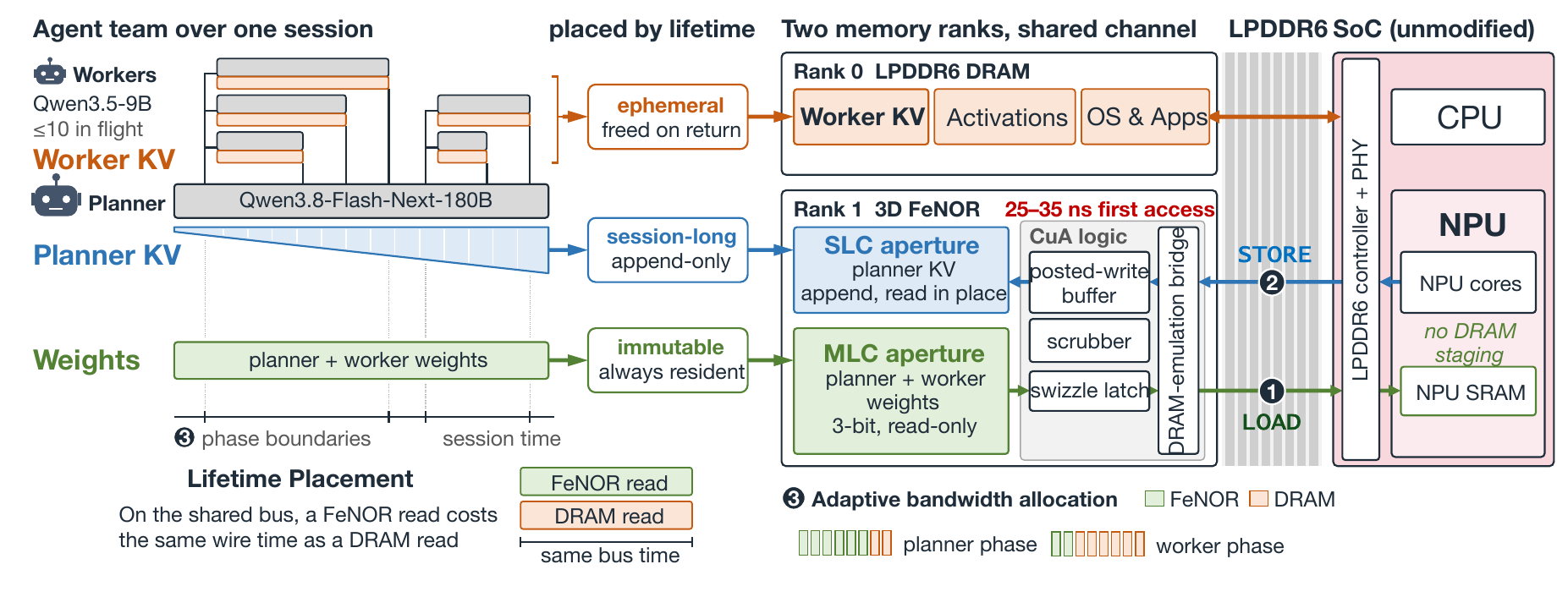}
\vspace{-14pt}
\caption{\sysname{} overview. FeNOR joins DRAM as Rank~1 on an unmodified 96-bit LPDDR6 bus, placing data by lifetime: \circled{1}~in-place MLC weight reads, \circled{2}~posted SLC planner KV appends, and \circled{3}~phase-adaptive subchannel splits.}
\label{fig:architecture}
\vspace{-6pt}
\end{figure*}
  
To an unmodified smartphone SoC, \sysname{} appears as a second LPDDR6 rank (Rank~1) sharing the 96-bit channel with LPDDR6 DRAM (Rank~0) via chip select. Its physical address space is partitioned into a read-only MLC aperture for co-resident model weights and a read/write SLC aperture for persistent planner KV caches, while Rank~0 DRAM hosts OS state, activations, and ephemeral worker KV caches. Fig.~\ref{fig:architecture} traces the three core runtime operations. \textbf{\circled{1}~Weight read:} the NPU issues standard DMA reads to the MLC aperture; the under-array bridge converts each \texttt{ACT} into staggered bank activations, the swizzle latch aligns 3-bit cells into byte-aligned weights, and bursts stream directly into NPU SRAM without DRAM staging (Sec.~\ref{sec:read_path}). \textbf{\circled{2}~Planner KV append:} planner KV lines are written via ordinary \texttt{STORE}s to the SLC aperture, absorbed at DRAM write timing by a posted-write buffer, drained into SLC banks in the background, and later re-read in place during attention (Sec.~\ref{sec:write_path}). \textbf{\circled{3}~Adaptive bandwidth allocation:} at each execution phase boundary, the runtime dynamically partitions the eight LPDDR6 subchannels between DRAM and FeNOR with work-conserving idle-channel borrowing (Sec.~\ref{sec:runtime}).

\section{\sysname{} Architecture}
\label{sec:architecture}

\subsection{Mutability- and Lifecycle-Tiered  Organization}
\label{sec:die}

\sysname{} tiers data by \emph{write mutability and agent lifecycle} rather than access hotness (Sec.~\ref{sec:data_classes}). Because DRAM and FeNOR share the same LPDDR6 bus, reading from either rank consumes identical wire time, whereas copying ``hot'' experts into DRAM wastes both bus bandwidth and energy. Each FeNOR die therefore splits its 3D array into two co-fabricated regions above the CuA logic layer (Fig.~\ref{fig:architecture}). The \textbf{MLC region} stores \emph{immutable} weights—including both the large planner and the lightweight worker models—programmed once at model installation and read in place during inference, eliminating model swapping across planner and worker phases. The \textbf{SLC region} stores \emph{persistent, append-only} planner KV caches, whose 100~ns single-cell writes and $>10^{11}$ endurance absorb long-horizon context growth without block erase or write amplification. Conversely, \textbf{LPDDR6 DRAM} holds \emph{mutable} state (activations and OS memory) alongside \emph{ephemeral worker KV caches}: because workers use a compact model and discard their context immediately upon completing a subtask, keeping their KV in DRAM avoids unnecessary flash writes and frees DRAM capacity as soon as each worker returns to the planner.
\begin{figure}[t]
\centering
\includegraphics[width=\columnwidth]{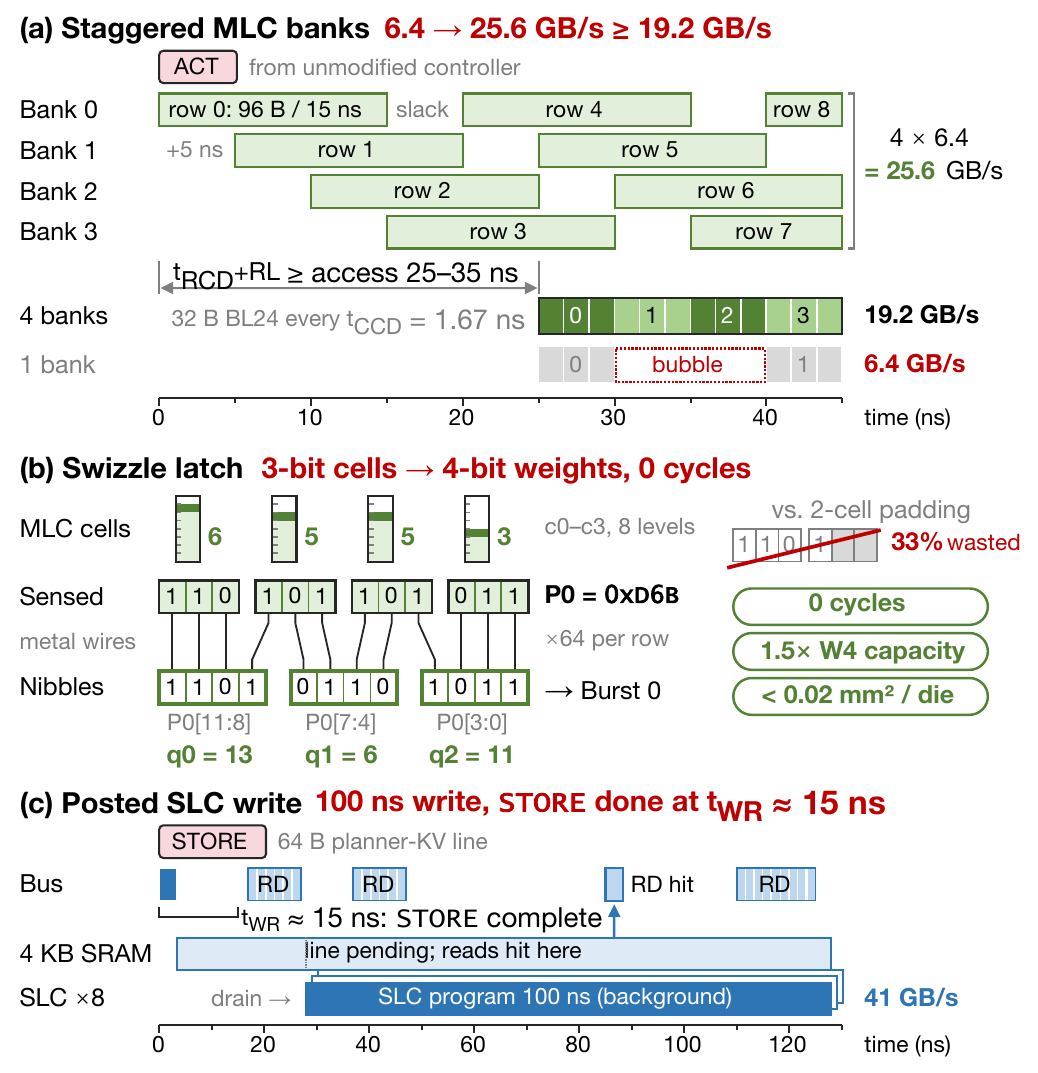}
\vspace{-14pt}
\caption{CuA datapath: (a)~staggered 4-bank MLC reads, (b)~zero-cycle W4 swizzle latch, and (c)~posted SLC writes.}
\label{fig:swizzle_pipeline}
\vspace{-6pt}
\end{figure}
Each NOR stack integrates $H\in\{4,6,8\}$ vertically stacked 68.15~mm$^2$ dies with $L\in\{16,32\}$ monolithic array layers per die over a single shared 96-bit LPDDR6 interface, scaling total stack capacity from 53.6~GB ($H\!=\!4, L\!=\!16$) to 221~GB ($H\!=\!8, L\!=\!32$), providing 49--197~GB of MLC weight capacity alongside 4--24~GB of SLC KV capacity. Across this design space, array depth $L$ sets per-die density within the sensing window while die count $H$ scales package capacity; partitioning the die area between MLC and SLC trades weight-capacity margin for larger on-package KV retention while comfortably fitting 100B--300B+ W4 models and co-resident worker models on an 8-die stack.

\noindent\textbf{Thermal-aware packaging.} Because 3-bit MLC threshold drift accelerates with temperature, placing the FeNOR stack atop an 85$^\circ$C SoC in a conventional package-on-package (PoP) layout degrades state retention. \sysname{} instead adopts a wafer-level multi-chip module (WMCM) layout that places the FeNOR stack side by side with the SoC and DRAM under a shared vapor chamber (Fig.~\ref{fig:package}), keeping the stack near 50$^\circ$C under sustained inference and slowing threshold drift by $3.5\times$. A lightweight CuA scrubber tracks thermal exposure and re-verifies drifted rows only while the phone is charging, so scrubbing never contends with inference.

\subsection{Read Path: DRAM-Emulation and Swizzle Latch}
\label{sec:read_path}

\noindent\textbf{Bus-aligned rows and DRAM-emulation bridge.} Each MLC bank senses $2^8=256$ cells per wordline (96~B, or three 32~B \texttt{BL24} bursts), holding 192 4-bit weights (W4) or 256 3-bit weights (W3) with a power-of-two column decoder and no leftover bits; each SLC bank senses a 64~B KV line (two bursts). Supported by LPDDR6's per-rank Mode Register (\texttt{MR}) timing configuration, the CuA bridge of each subchannel emulates deterministic DRAM read timing across the full stack capacity (addressed via upper row-address bits). Upon receiving \texttt{ACT}, the bridge triggers four MLC banks staggered by 5.0~ns to fill a rolling row buffer within the programmed $t_{RCD}+RL$ window (Eq.~\ref{eq:rank_latency}), and subsequently streams 32~B bursts every $t_{CCD}=1.67$~ns on \texttt{RD} commands. Because FeNOR reads are non-destructive, \texttt{PRE} and \texttt{REF} commands update only bridge state. Four staggered banks provide 25.6~GB/s per subchannel—exceeding the 19.2~GB/s subchannel wire rate and providing slack against bank-cycle variations—to satisfy Eq.~\ref{eq:rank_bw} (Fig.~\ref{fig:swizzle_pipeline}(a)).

\noindent\textbf{Swizzle latch.} Storing 4-bit weights in 3-bit MLC cells via naive two-cell padding wastes 33\% of capacity, while software bit-unpacking stalls the NPU vector pipeline. The swizzle latch eliminates both overheads by exploiting the natural 12-bit alignment between LPDDR6 subchannels, four 3-bit cells, and three 4-bit weights. At model installation, every three 4-bit weights (12~bits) are striped across four 3-bit cells. During a read, the 256 sensed cells form 64 four-cell tuples; for each tuple $m$, fixed metal routing between the sense amplifiers and the burst serializer directly wires the 12 sensed bits $P_m$ into three consecutive 4-bit weight nibbles $(P_m[11\!:\!8],\, P_m[7\!:\!4],\, P_m[3\!:\!0])$ on the byte bus (Fig.~\ref{fig:swizzle_pipeline}(b)), while a 2:1 mux supports native W3 mode. Because reassembly is pure wiring, the latch adds zero cycle latency and $<$0.02~mm$^2$ per die while increasing effective W4 capacity by $1.5\times$ over padding. Quantization group scales are stored in a separate byte-aligned table ($<$1.6\% metadata traffic).

\subsection{Write Path: Posted SLC Writes}
\label{sec:write_path}

Although DG SLC switches cells in place in 100~ns without block erase, it remains slower than LPDDR6 write timing ($t_{WR}\approx 15$~ns). Each subchannel bridge therefore integrates a 4~KiB SRAM posted-write buffer ($<$0.08~mm$^2$ per die) that accepts 64~B planner KV lines at full bus speed and drains them in the background across eight interleaved SLC banks (Fig.~\ref{fig:swizzle_pipeline}(c)). Together, eight banks per subchannel sustain 41~GB/s of aggregate write bandwidth across the 96-bit interface—more than $20\times$ the peak KV append rate during inference (Sec.~\ref{sec:eval_write})—preventing buffer overflow. Read requests that hit pending lines in the buffer are served directly from SRAM, and the bridge schedules bank drains between attention read bursts (or steers writes to idle banks) so background SLC programming never blocks latency-critical reads.

\subsection{Runtime and Adaptive Bandwidth Allocation}
\label{sec:runtime}
\begin{figure}[t]
\centering
\includegraphics[width=\columnwidth]{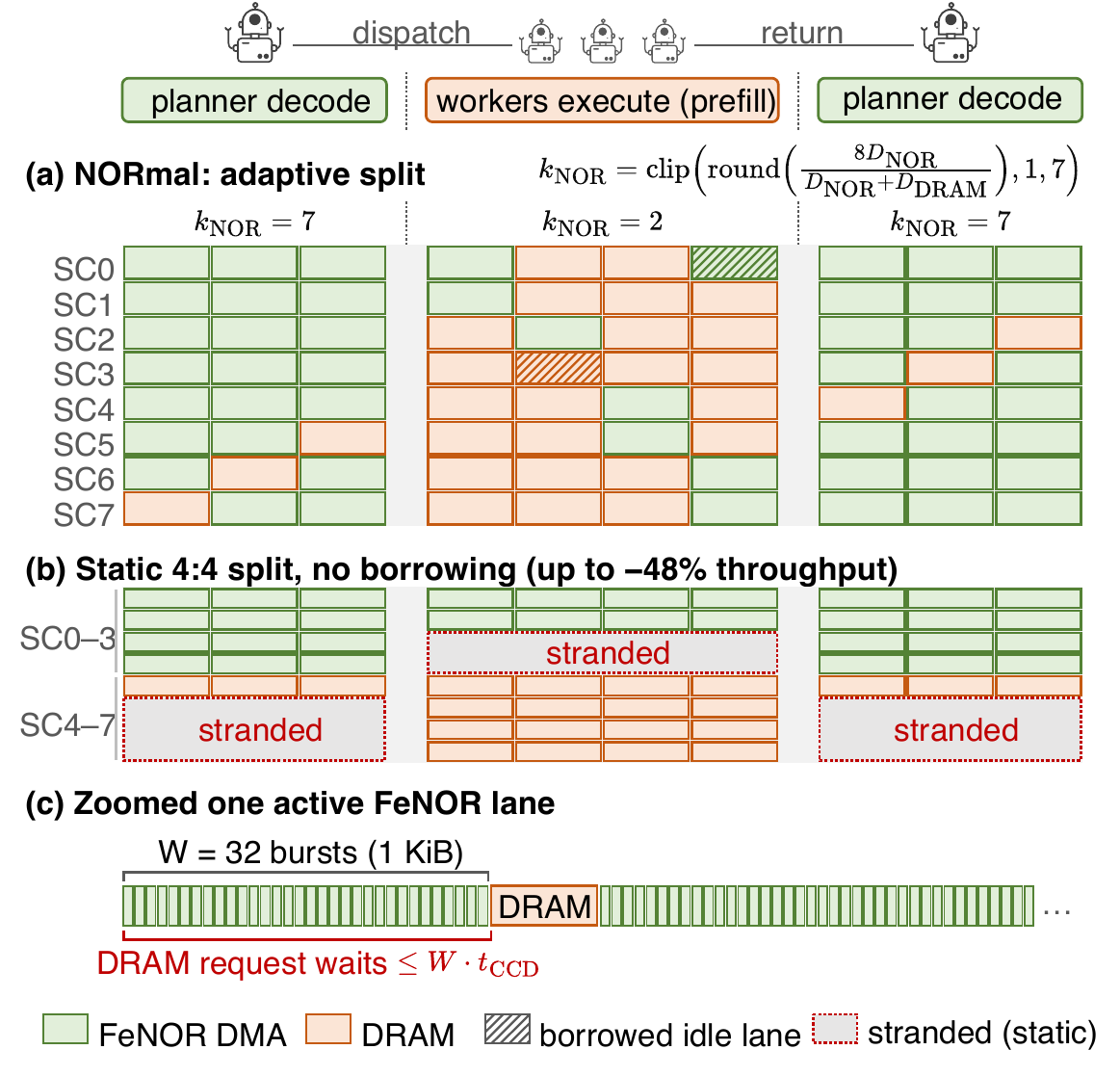}
\vspace{-4pt}
\caption{LPDDR6 subchannel sharing: (a)~phase-adaptive split, (b)~static 4:4 split, and (c)~bounded DRAM preemption.}
\label{fig:bus_alloc}
\vspace{-6pt}
\end{figure}

\noindent\textbf{Lifecycle-aware placement.} At model load time, the runtime maps each data class to its matching tier without modifying model code. Co-resident planner and worker model weights reside permanently in FeNOR MLC and stream directly into NPU SRAM via DMA. Planner KV caches are appended to FeNOR SLC and read in place during attention, whereas ephemeral worker KV caches remain in LPDDR6 DRAM and are reclaimed immediately when a worker completes its subtask. When an extreme single-agent session on a KV-heavy dense model exceeds the default  SLC aperture, which can be enlarged beyond the default by trading MLC area for SLC capacity, no workers are active in DRAM, so the runtime spills the remaining KV tail into the usable DRAM budget before ever touching UFS, absorbing the full session on package.

\noindent\textbf{Adaptive bandwidth allocation.} Because DRAM and FeNOR share the eight 12-bit LPDDR6 subchannels, static subchannel partitioning leaves bandwidth stranded as execution shifts between FeNOR-heavy planner weight/KV streaming and DRAM-heavy worker execution or prefill (Fig.~\ref{fig:bus_alloc}). Since weights and KV lines are striped across all eight subchannel address slices, the NPU runtime steers subchannel bandwidth in software without modifying the SoC memory controller. At each layer or agent-stage boundary, the runtime estimates the pending FeNOR traffic $D_{\text{NOR}}$ (weights, routed experts, and SLC KV reads) and DRAM traffic $D_{\text{DRAM}}$ (activations, worker KV, and reserved OS traffic), and dispatches concurrent FeNOR DMA descriptors to
\begin{equation}
k_{\text{NOR}} = \text{clip}\!\left(\!\text{round}\!\left(\frac{8\, D_{\text{NOR}}}{D_{\text{NOR}} + D_{\text{DRAM}}}\right)\!,\, 1,\, 7\!\right)
\label{eq:k_nor}
\end{equation}
active subchannel slices, rotating the active set across micro-batches so every address slice drains evenly while leaving at least one subchannel queue free of FeNOR DMA traffic for latency-critical DRAM requests. Crucially, the runtime pairs this phase-aware split with \emph{work-conserving idle-subchannel borrowing}: whenever either rank has no pending descriptors for its assigned slices within a scheduling window, the runtime steers ready DMA descriptors from the other rank into those idle subchannel queues rather than leaving bus wires unused. Finally, the runtime groups FeNOR DMA accesses on each active subchannel into contiguous batches of $W=32$ bursts (1~KiB); combined with the unmodified controller's standard command batching and LPDDR6's fast dynamic \texttt{NT-ODT} switching, this amortizes rank-switch turnaround while bounding high-priority DRAM preemption delay to at most $W\cdot t_{CCD}=53.3$~ns.

\section{Evaluation}
\label{sec:evaluation}

\begin{figure*}[t]
\centering
\includegraphics[width=\textwidth]{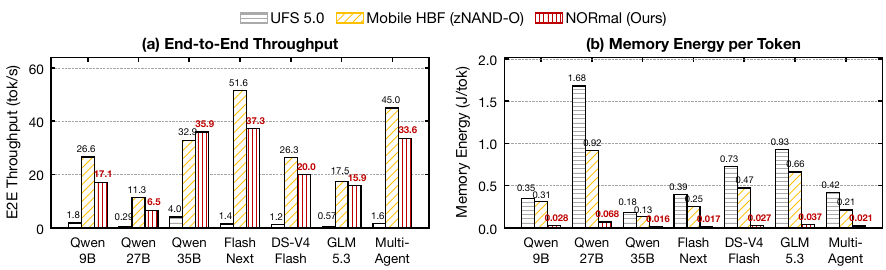}
\vspace{-16pt}
\caption{End-to-end performance across seven workloads: (a)~pooled output throughput (\texttt{tok/s}, excluding trace idle time) and (b)~memory subsystem energy per generated token (\texttt{J/tok}).}
\label{fig:e2e_eval}
\vspace{-8pt}
\end{figure*}

\subsection{Methodology}
\label{sec:methodology}

\noindent\textbf{Simulator and hardware configuration.} We evaluate \sysname{} using a full-stack serving simulator built upon TokenSim~\cite{tokensim2025,tokensimv22026}\footnote{\url{https://github.com/pku-lemonade/TokenSim}}, modeling Xiaomi XRING O3, a flagship mobile SoC equipped with a 200~TOPS INT4/INT8 NPU and a 96-bit LPDDR6 memory subsystem. All configurations provision 16~GB of LPDDR6 DRAM, with 8~GiB pinned by the mobile OS and background applications.
Of the remaining 8~GiB, about 0.5~GiB is reserved for the inference framework and runtime, leaving 7.46~GiB for weights, KV caches, and activations.
FeNOR array timing and energy follow fabricated devices~\cite{igzofefet2025,pcaa2025} (Table~\ref{tab:device_comparison}) and assume a passive mobile thermal envelope. \sysname{} uses an 8-die, 32-layer FeNOR stack sharing the 96-bit LPDDR6 channel as Rank~1 (197~GB MLC for weights and up to 24~GB SLC for persistent KV caches).

\noindent\textbf{Workloads and agent traces.} Table~\ref{tab:workloads} summarizes the seven evaluated workloads: two dense models, four sparse MoE models spanning 35B to 314B parameters, and a \textbf{Multi-Agent} planner--worker team that pairs a \textbf{Qwen3.8-Flash-Next} planner with four concurrent \textbf{Qwen3.5-9B} workers; we use same-family models so that planner and workers share a tokenizer and prompt format (Sec.~\ref{sec:workloads}). To capture realistic long-horizon agent execution, we replay 215 intact single-agent coding sessions up to 962K context tokens and 164 multi-agent sessions with up to 1,970 turns and four-way parallel worker delegation from production coding-agent traces~\cite{wekatraces2026}\footnote{\url{https://huggingface.co/datasets/semianalysisai/cc-traces-weka-062126}}. Weights use group-quantized W4 (packed via the swizzle latch; model sizes in Table~\ref{tab:workloads} include group-quantization scales and non-expert layers) and KV caches use INT8 with architecture-native recurrent state formats.

\begin{table}[t]
\centering
\caption{Evaluated Hardware, Agent Traces, and Models.}
\label{tab:workloads}
\renewcommand{\arraystretch}{1.08}
\fontsize{7}{8.2}\selectfont
\setlength{\tabcolsep}{1.5pt}
\begin{tabular}{@{}>{\raggedright\arraybackslash}p{\dimexpr1\columnwidth-0pt\relax}@{}}
\toprule
\textbf{Platform:} Xiaomi XRING O3 Mobile SoC, 200~TOPS NPU, W4A8, INT8 KV \\
\hspace{0.8em}LPDDR6: 96-bit ($8\times 12$-b subch.), 153.6~GB/s \quad Host DRAM: 16~GB (8~GiB avail.) \\
\hspace{0.8em}NOR: 8-Hi$\times$32L CuA, 96~B MLC / 64~B SLC, $4\times$/$8\times$ bank interleaving \\
\bottomrule
\end{tabular}
\par\vspace{2pt}
\begin{tabular}{@{}>{\raggedright\arraybackslash}p{\dimexpr0.14\columnwidth-2.4pt\relax}>{\raggedright\arraybackslash}p{\dimexpr0.23\columnwidth-2.4pt\relax}>{\raggedright\arraybackslash}p{\dimexpr0.22\columnwidth-2.4pt\relax}>{\raggedright\arraybackslash}p{\dimexpr0.18\columnwidth-2.4pt\relax}>{\raggedright\arraybackslash}p{\dimexpr0.23\columnwidth-2.4pt\relax}@{}}
\toprule
\textbf{Mode} & \textbf{Tier org.} & \textbf{BW (DRAM/\allowbreak Flash)} & \textbf{Cap. (DRAM/\allowbreak Flash)} & \textbf{KV policy} \\
UFS 5.0 & DRAM + Serial NAND & 153.6 / 10.8~GB/s & 8~GiB / 512~GB & LRU, UFS swap \\
Mobile HBF & DRAM + $3\times$4-Hi zNAND-O & 153.6 / 200~GB/s & 8~GiB / 192~GB & LRU, UFS spill / recompute \\
\sysname{} & Shared LPDDR6 Rank~1 & Shared 153.6~GB/s & 8~GiB / 221~GB & SLC/DRAM, Lifecycle-tiered \\
Ideal LPDDR6 & Unlimited Rank~0 DRAM & 153.6 GB/s / — & $\infty$ / 0~GB & In-DRAM full retention \\
\bottomrule
\end{tabular}
\par\vspace{2pt}
\begin{tabular}{@{}>{\raggedright\arraybackslash}p{\dimexpr0.18\columnwidth-2.4pt\relax}>{\raggedright\arraybackslash}p{\dimexpr0.52\columnwidth-2.4pt\relax}>{\raggedright\arraybackslash}p{\dimexpr0.1\columnwidth-2.4pt\relax}>{\raggedright\arraybackslash}p{\dimexpr0.1\columnwidth-2.4pt\relax}>{\raggedright\arraybackslash}p{\dimexpr0.1\columnwidth-2.4pt\relax}@{}}
\toprule
\textbf{Trace} & \textbf{Type (Sessions, Max Context)} & \textbf{Reqs} & $\boldsymbol{\bar{P}}$ & $\boldsymbol{\bar{G}}$ \\
\texttt{single} & Single-agent (215 sessions, $\le$962K) & 17\,122 & 223\,397 & 1\,159 \\
\texttt{long} & Long-context (62 sessions, $>$256K) & 10\,276 & 297\,266 & 1\,309 \\
\texttt{multi-agent} & 1~Planner + 4~Workers (164 sessions) & 54\,910 & 94\,757 & 815 \\
\bottomrule
\end{tabular}
\par\vspace{2pt}
\begin{tabular}{@{}>{\raggedright\arraybackslash}p{\dimexpr0.26\columnwidth-2.5pt\relax}>{\raggedright\arraybackslash}p{\dimexpr0.10\columnwidth-2.5pt\relax}>{\raggedright\arraybackslash}p{\dimexpr0.17\columnwidth-2.5pt\relax}>{\raggedright\arraybackslash}p{\dimexpr0.17\columnwidth-2.5pt\relax}>{\raggedright\arraybackslash}p{\dimexpr0.15\columnwidth-2.5pt\relax}>{\raggedright\arraybackslash}p{\dimexpr0.15\columnwidth-2.5pt\relax}@{}}
\toprule
\textbf{Model} & \textbf{Layers} & \textbf{Experts (Top-$k$)} & \textbf{Params (Act.)} & \textbf{Model Size} & \textbf{KV / tok} \\
Qwen3.5-9B & 32 & Dense (—) & 9B (8B) & 4.43~GiB & 17.00~KiB \\
Qwen3.8-27B & 64 & Dense (—) & 27B (26B) & 13.31~GiB & 34.00~KiB \\
Qwen3.6-35B-A3B & 40 & 256 (Top-8) & 35B (3B) & 17.15~GiB & 10.63~KiB \\
Qwen3.8-Flash-Next & 48 & 512 (Top-10) & 177B (7B) & 87.84~GiB & 13.15~KiB \\
DeepSeek-V4-Flash & 43 & 256 (Top-6) & 285B (13B) & 140.69~GiB & 3.57~KiB \\
GLM-5.3-Flash & 45 & 288 (Top-8) & 314B (17B) & 155.02~GiB & 6.21~KiB \\
\bottomrule
\end{tabular}
\end{table}

\noindent\textbf{Baselines.} We compare three mobile memory systems alongside an idealized reference: \textbf{(1)~UFS~5.0}~\cite{powerinfer2}, which pairs LPDDR6 DRAM with 10.8~GB/s UFS~5.0 flash storage and demand-fetches uncached weights and spilled KV blocks; \textbf{(2)~Mobile HBF (zNAND-O)}~\cite{samsung2025znando,flashaccel2026}, which integrates a 3-stack mobile HBF tier providing a dedicated 200~GB/s read-mostly weight link ($10^5$ P/E endurance, 4~KiB page access) while keeping active KV caches in LPDDR6; \textbf{(3)~\sysname{} (Ours)}, which attaches the dual-tier MLC/SLC FeNOR stack directly to the shared 96-bit LPDDR6 bus with adaptive subchannel allocation; and \textbf{(4)~Ideal LPDDR6}, an upper-bound reference with unlimited LPDDR6 DRAM capacity.

\subsection{End-to-End Throughput, Latency, and Energy}
\label{sec:eval_e2e}

\noindent\textbf{End-to-end throughput and energy.} Fig.~\ref{fig:e2e_eval}(a) compares pooled end-to-end throughput across all seven workloads. Because UFS~5.0 is bottlenecked by its 10.8~GB/s serial link and tens-of-microseconds page latency during weight streaming and KV swapping, it sustains only 0.29--4.05~tok/s. By executing weights in place from MLC and retaining long-horizon KV caches in SLC at LPDDR6 bus speed, \sysname{} achieves 6.50--37.26~tok/s, delivering an \textbf{$8.9\times$--$28.0\times$ end-to-end speedup over UFS~5.0} ($26.1\times$ on \textbf{Flash Next}, $28.0\times$ on \textbf{GLM~5.3}, and $21.4\times$ on \textbf{Multi-Agent}). Against Mobile HBF, the bandwidth tradeoff of Sec.~\ref{sec:device_bg} is visible. On six of the seven workloads, HBF's dedicated 200~GB/s link gives it $1.10\times$--$1.73\times$ higher throughput, and \sysname{} reaches 58\%--91\% of HBF. Decode on these sessions is bound by weight streaming, and HBF adds a link while \sysname{} shares the LPDDR6 bus. On \textbf{Qwen~35B}, \sysname{} outperforms Mobile HBF (35.9 vs.\ 32.9~tok/s) even on HBF's supported session subset, because HBF repeatedly recomputes prefixes when inactive session caches are evicted from DRAM. Sec.~\ref{sec:eval_tasks} shows that the ranking reverses once KV caches exceed DRAM. Fig.~\ref{fig:e2e_eval}(b) shows that \sysname{} reduces memory subsystem energy to \textbf{0.016--0.068~J/tok}, an \textbf{$11.3\times$--$27.0\times$ reduction over UFS~5.0} and an \textbf{$8.1\times$--$17.8\times$ reduction over Mobile HBF} (0.13--0.92~J/tok). Mobile HBF's high energy stems from activating wide 4~KiB NAND pages and waking multi-TSV HBF dies, whereas \sysname{} senses only exact 96~B/64~B lines over low-capacitance CuA routing.

\begin{figure}[t]
\centering
\includegraphics[width=\columnwidth]{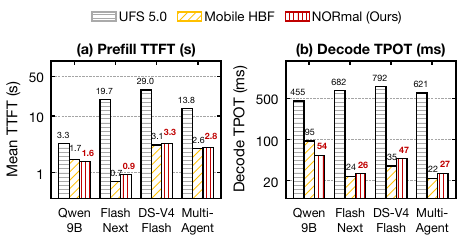}
\vspace{-16pt}
\caption{Latency breakdown under identical KV retention: (a)~mean prefill TTFT and (b)~mean decode TPOT.}
\label{fig:ttft_tpot}
\vspace{-8pt}
\end{figure}

\noindent\textbf{Prefill TTFT and decode TPOT.} Fig.~\ref{fig:ttft_tpot} breaks down latency into mean prefill TTFT and decode TPOT under an identical KV retention policy across systems. During prefill (Fig.~\ref{fig:ttft_tpot}(a)), UFS~5.0 incurs 3.3--29.0~s of TTFT because cold model weights and historical KV prefixes must stream across the storage link. \sysname{} cuts mean TTFT to 0.9--3.3~s ($2.1\times$--$21.6\times$ faster than UFS~5.0). On the KV-intensive dense model (\textbf{Qwen~9B}: 1.6~s vs.\ 1.7~s), \sysname{} has slightly lower TTFT than Mobile HBF, which must spill or rebuild evicted prefixes. HBF is 0.1--0.2~s faster on the three other workloads. During decode (Fig.~\ref{fig:ttft_tpot}(b)), \sysname{} reduces TPOT from 455--792~ms on UFS~5.0 to 26--54~ms ($8.4\times$--$26.1\times$ lower). It cuts \textbf{Qwen~9B} TPOT by $1.75\times$ relative to Mobile HBF (54~ms vs.\ 95~ms). On the MoE and multi-agent workloads, HBF's extra link gives it $1.1\times$--$1.3\times$ lower TPOT.

\subsection{Task Finish Time, Long-Horizon Scaling, and Multi-Agent Gain}
\label{sec:eval_tasks}

\begin{figure}[t]
\centering
\includegraphics[width=\columnwidth]{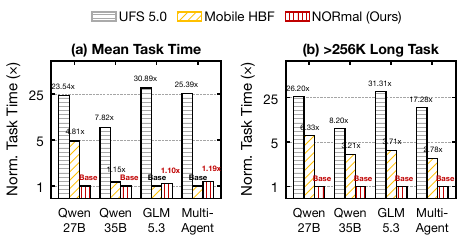}
\vspace{-16pt}
\caption{Normalized task finish time (\texttt{Base} = $1.00\times$): (a)~all sessions and (b)~long-horizon sessions ($>$256K tokens).}
\label{fig:task_finish}
\vspace{-8pt}
\end{figure}

\begin{figure}[t]
\centering
\includegraphics[width=\columnwidth]{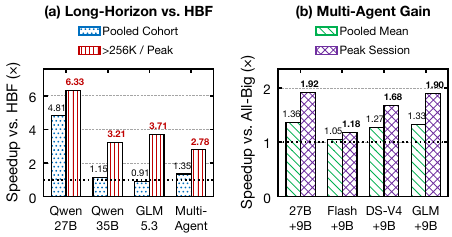}
\vspace{-16pt}
\caption{(a)~\sysname{} speedup over Mobile HBF on pooled vs.\ $>$256K sessions, and (b)~team speedup over \emph{Planner-Only}.}
\label{fig:hbf_comparison}
\vspace{-8pt}
\end{figure}

\noindent\textbf{Session-level task finish time.} Without UFS KV spilling, a read-only weight tier fails once a session's KV footprint exceeds usable DRAM: Mobile HBF fails capacity checks on 87 of 215 sessions (40.5\%) for \textbf{Qwen~27B} and 20 of 215 sessions (9.3\%) for \textbf{Qwen~9B}. With UFS spilling and prefix recomputation enabled so every session in the cohort runs to completion, Fig.~\ref{fig:task_finish} reports normalized task finish time relative to the fastest system (\texttt{Base}). Across the full cohort (Fig.~\ref{fig:task_finish}(a)), \sysname{} is the fastest system (\texttt{Base}) on both \textbf{Qwen~27B} (where Mobile HBF takes $4.81\times$ longer and UFS~5.0 takes $23.54\times$ longer) and \textbf{Qwen~35B} (where HBF takes $1.15\times$ longer), while remaining within $1.10\times$--$1.19\times$ of HBF's dedicated 200~GB/s link on \textbf{GLM~5.3} and \textbf{Multi-Agent}. On the 62 long-horizon tasks exceeding 256K context tokens (256K--962K, Fig.~\ref{fig:task_finish}(b)), \textbf{\sysname{} is \texttt{Base} across all four workloads}: Mobile HBF takes \textbf{$2.78\times$--$6.33\times$ longer} ($6.33\times$ on \textbf{Qwen~27B}, $3.21\times$ on \textbf{Qwen~35B}, $3.71\times$ on \textbf{GLM~5.3}, and $2.78\times$ on \textbf{Multi-Agent}), and UFS~5.0 takes $8.20\times$--$31.31\times$ longer.

\noindent\textbf{Why \sysname{} beats HBF on long horizons.} Fig.~\ref{fig:hbf_comparison}(a) contrasts \sysname{}'s speedup over Mobile HBF between the pooled cohort and $>256\text{K}$/peak sessions. Once active and retained KV caches exceed the 8~GiB DRAM budget, Mobile HBF cannot absorb KV writes on package within a phone's passive thermal and endurance budgets; it must either spill KV to UFS~5.0 (incurring 47.6~TB of cumulative UFS reads on \textbf{Qwen~27B}) or evict and recompute up to 499M prefix tokens on \textbf{GLM~5.3}. \sysname{} absorbs up to 23.6~GB of persistent KV caches in byte-addressable FeNOR SLC (plus up to 8~GiB in usable DRAM on extreme single-agent sessions) and reads SLC KV in place without promotion to DRAM, reaching up to $11.98\times$ single-session speedup on \textbf{Qwen~27B} and $8.08\times$ on \textbf{Qwen~35B}.

\noindent\textbf{Multi-agent cooperation.} Fig.~\ref{fig:hbf_comparison}(b) evaluates a team of one planner and four workers. Beyond the default \textbf{Flash~Next} planner, it also varies the planner (\textbf{Qwen~27B}, \textbf{DS-V4~Flash}, or \textbf{GLM~5.3}) while keeping the \textbf{Qwen~9B} worker model in FeNOR MLC. Because both models remain permanently resident in MLC while the four concurrent workers allocate their ephemeral KV caches in LPDDR6 DRAM, \sysname{} eliminates the tens of gigabytes of UFS model-swapping traffic required by conventional mobile runtimes and achieves a \textbf{$1.05\times$--$1.36\times$ pooled speedup} and \textbf{$1.18\times$--$1.92\times$ peak session speedup} over running all turns on the planner model (\emph{Planner-Only}).

\subsection{Adaptive Bandwidth Allocation and Device Endurance}
\label{sec:eval_ablation}
\label{sec:eval_write}

\begin{figure}[t]
\centering
\includegraphics[width=\columnwidth]{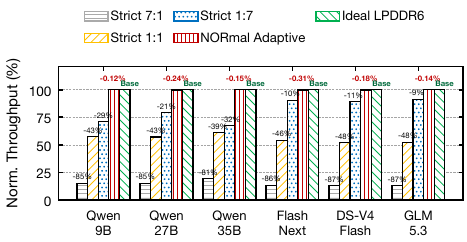}
\vspace{-16pt}
\caption{Throughput under static subchannel splits versus \sysname{} adaptive allocation, normalized to \texttt{Ideal LPDDR6}.}
\label{fig:bandwidth_ablation}
\vspace{-8pt}
\end{figure}

\begin{figure}[t]
\centering
\includegraphics[width=\columnwidth]{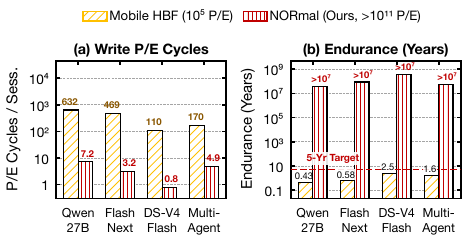}
\vspace{-16pt}
\caption{Endurance on the 962K-context session: (a)~flash P/E cycles per session and (b)~projected lifetime in years.}
\label{fig:endurance}
\vspace{-8pt}
\end{figure}

\noindent\textbf{Adaptive bus sharing vs.\ static splits.} Fig.~\ref{fig:bandwidth_ablation} compares \sysname{}'s adaptive subchannel allocator against static DRAM-to-FeNOR subchannel reservations (\texttt{Strict 7:1}, \texttt{1:1}, and \texttt{1:7} without idle borrowing), normalized to \texttt{Ideal LPDDR6} (\texttt{Base}). Static \texttt{7:1} starves FeNOR weight streaming and loses 81\%--87\% of throughput, while static \texttt{1:1} loses 39\%--48\%. Even a FeNOR-heavy \texttt{Strict 1:7} split loses 9\%--32\% of throughput (losing 29\% on \textbf{Qwen~9B} and 32\% on \textbf{Qwen~35B}) because a single DRAM subchannel throttles DRAM-resident recurrent state and KV traffic. By dynamically matching subchannel shares to phase demand (Eq.~\ref{eq:k_nor}) with work-conserving idle borrowing, \sysname{} incurs only \textbf{0.12\%--0.31\% throughput loss ($>99.6\%$ of \texttt{Ideal LPDDR6})} across all six models. Furthermore, on the read path, four-bank interleaving with the zero-cycle swizzle latch eliminates the $5.4\times$--$11.3\times$ 4~KiB page read amplification of NAND tiers and expands effective W4 weight capacity by $1.5\times$ over two-cell padding, enabling the 155.02~GiB \textbf{GLM~5.3} model (Table~\ref{tab:workloads}) to fit within a single 8-die stack.

\noindent\textbf{Write volume and multi-year device endurance.} Finally, Fig.~\ref{fig:endurance} evaluates flash write endurance on the longest multi-turn coding session (962K context, 794 requests, 812K generated tokens; and 1,970 requests for \textbf{Multi-Agent}). Because a phone's passive thermal budget forbids sustained multi-watt 100~pJ/bit NAND programming during normal operation, we evaluate a relaxed upper bound for Mobile HBF that ignores thermal limits, permits KV offloading into HBF, and wear-levels writes across the entire post-weight slack capacity of the 192~GB HBF package (177.7~GB on \textbf{Qwen~27B}, 97.7~GB on \textbf{Flash~Next}, 40.9~GB on \textbf{DS-V4~Flash}, and 20.8~GB on \textbf{Multi-Agent}). Even with buffered append-only writes~\cite{flashaccel2026}, 4~KiB page granularity and repeated KV eviction and promotion between DRAM and flash write 3.5--112.2~TB to flash in a single session, consuming \textbf{110--632 P/E cycles per session} across the slack capacity (Fig.~\ref{fig:endurance}(a)). Under one such heavy agent session per day, Mobile HBF ($10^5$ P/E limit) still wears out in just \textbf{0.43--2.48 years} (5.2~months on \textbf{Qwen~27B} and 7.0~months on \textbf{Flash~Next}), falling well short of the 5-year smartphone lifespan requirement even under this generous wear-leveling assumption (Fig.~\ref{fig:endurance}(b)). In contrast, because \sysname{} writes only newly appended 64~B KV lines in place to its 23.6~GB SLC region without block erase and reads historical SLC KV blocks directly in place without promoting them back to DRAM, it writes only 18.6--170.3~GB per session (\textbf{0.8--7.2 P/E cycles}, a $30\times$--$659\times$ reduction in physical write volume). Combined with FeNOR's $>10^{11}$ P/E endurance, \sysname{} comfortably exceeds the 5-year smartphone target ($>10^7$~years), eliminating write wear-out as a deployment barrier for on-device agents.

\section{Conclusion}
\label{sec:conclusion}

We presented \sysname{}, the first main-memory architecture for 3D ferroelectric NOR flash. It turns flash from block storage into byte-addressable inference memory by attaching our fabricated dual-tier 3D FeNOR as a native second rank on an unmodified mobile LPDDR6 interface. Our characterization shows that on-device large-model agents need memory-speed weight access from a capacity tier, writable capacity for KV beyond DRAM, and lifecycle-aware KV placement. Because FeNOR meets the latency bound of an LPDDR rank, a deterministic DRAM-emulation bridge with four-bank interleaving, a zero-cycle W4/W3 swizzle latch, a lifecycle-tiered MLC/SLC die with posted SLC writes, and an adaptive bus-sharing runtime are enough to host 200B-class models and long-horizon agent KV caches on package. 

\bibliographystyle{ACM-Reference-Format}
\bibliography{references}

\end{document}